\pdfoutput=1
\documentclass[
    reprint, twocolumn,
    aps,prx,10pt,amsmath,amssymb,
    longbibliography,superscriptaddress,
    showpacs,preprintnumbers,
    groupedaddress]{revtex4-2}

\usepackage{xr}
\makeatletter
\newcommand*{\addFileDependency}[1]{
  \typeout{(#1)}
  \@addtofilelist{#1}
  \IfFileExists{#1}{}{\typeout{No file #1.}}
}
\makeatother

\newcommand*{\myexternaldocument}[1]{
    \externaldocument{#1}
    \addFileDependency{#1.tex}
    \addFileDependency{#1.aux}
}

\myexternaldocument{supp}

\usepackage[colorlinks,bookmarks=false,citecolor=blue,linkcolor=blue,urlcolor=blue]{hyperref}
\usepackage{oubraces}
\usepackage{bm,bbm,mathbbol}
\usepackage{physics}
\usepackage{mathtools}
\usepackage{amsmath, amsthm, amssymb, amsfonts}
\usepackage{bbold}
\usepackage{esvect}
\usepackage[dvipsnames]{xcolor}

\usepackage{epsfig}
\usepackage{array}
\usepackage{multirow}
\usepackage{graphicx}
\usepackage[normalem]{ulem}

\usepackage{float}
\usepackage[section]{placeins}
\usepackage{afterpage}

\usepackage{lipsum}

\usepackage{xifthen}

\makeatletter
\newcommand\footnoteref[1]{\protected@xdef\@thefnmark{\ref{#1}}\@footnotemark}
\makeatother

\newcommand{\myeqref}[1]{Eq.~\eqref{#1}}
\newcommand{\Eq}[1]{Eq.~\eqref{#1}}

\newcommand{\mb}[1]{{\boldsymbol{\mathbf{#1}}}}

\newcommand{\df}[0]{\delta\! f}

\newcommand{\mcO}[0]{{\mathcal{O}}}

\begin{document}

\title{Nonlinear Hall effect from long-lived valley-polarizing relaxons}

\author{Jae-Mo Lihm}
\email{jaemo.lihm@gmail.com}
\author{Cheol-Hwan Park}
\email{cheolhwan@snu.ac.kr}
\affiliation{Department of Physics and Astronomy, Seoul National University, Seoul 08826, Korea,}
\affiliation{Center for Correlated Electron Systems, Institute for Basic Science, Seoul 08826, Korea,}
\affiliation{Center for Theoretical Physics, Seoul National University, Seoul 08826, Korea}

\date{\today}

\begin{abstract}
The nonlinear Hall effect has attracted much attention due to the famous, widely adopted interpretation in terms of the Berry curvature dipole in momentum space.
Using \textit{ab initio} Boltzmann transport equations, we find a 60\% enhancement in the nonlinear Hall effect of $p$-doped GeTe and its noticeable frequency dependence, qualitatively different from the predictions based on the Berry curvature dipole.
The origin of these differences is long-lived valley polarization in the electron distribution arising from electron-phonon scattering.
Our findings await immediate experimental confirmation.
\end{abstract}

\footnotetext[1]{See Supplemental Material at
\url{http://link.aps.org/
supplemental/10.1103/PhysRevLett.132.106402}
which includes Refs.~\cite{2015ZhouDrag,2020ProtikDrag,2010XiaoRMP,2018TsirkinGyrotropic,2015GunnarssonFlucDiag,2013IsbergValley,2021MaliyovHighField,2017GiannozziQE,2013HamannONCVPSP,2018VanSettenPseudoDojo,1996PerdewPBE,2016KrempaskyGeTe,2020PizziWannier90,2017BezansonJulia,JuliaIterativeSolvers,2022LihmSpinPhoto,2006WangAHC,2022MachedaDoping,2019RoyoQuadrupole,2020BruninQuadrupole,2020JhalaniQuadrupole,JuliaBrillouin,JuliaInterpolations,2019AskarpourGeTe},
for the details of the BTE and relaxon formalism, additional computational results, and computational details.}

\newcommand{\citeSupp}[0]{Note1}

\maketitle

The nonlinear Hall (NLH) effect~\cite{2009DeyoBCD,2015SodemannPRL,2010Moore,2019MaNLH,2019KangNLH,2021DuReview,2021DuVertex,Sinha2022} is a nonlinear analogue of the Hall effect that describes a transverse current response to two electric fields.
Unlike its linear counterpart, the NLH effect occurs even in non-magnetic systems without an external magnetic field.
The NLH effect is attracting much attention due to its close connection to an intrinsic geometric property of the electronic structure~\cite{2015SodemannPRL,2019MatsyshynPRL}.
Applications such as probing electronic topology~\cite{2018FacioBCD}, radio-frequency rectification~\cite{2021KumarNLH}, and terahertz photodetection~\cite{2020IsobeRectification,2020GuoNLH,2021ZhangPNAS} are under active investigation.

\begin{figure}[b]
\centering
\includegraphics[width=1.0\columnwidth]{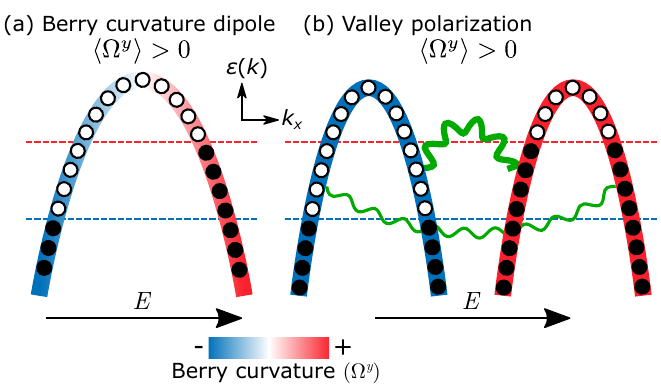}
\caption{
Mechanisms of the intrinsic NLH effect.
(a) Berry curvature dipole, and
(b) scattering-induced valley polarization.
The thickness of the wiggles represents the strength of the scattering.
}
\label{fig:schematic}
\end{figure}

The standard interpretation of the intrinsic NLH effect attributes the effect to the momentum-space dipole of the Berry curvature~\cite{2009DeyoBCD,2015SodemannPRL,2019MatsyshynPRL} (Fig.~\ref{fig:schematic}(a)).
This ``Berry curvature dipole'' picture underlies many experimental~\cite{2019KangNLH,2019MaNLH,Sinha2022} and theoretical~\cite{2010Moore,2021ZhangPNAS,2018FacioBCD} studies of the NLH effect.
An assumption key to this interpretation is the constant relaxation time approximation (CRTA).
In this approximation, both the NLH conductivity and the linear conductivity are proportional to the relaxation time, a phenomenological constant, with the prefactors being the Berry curvature dipole and the Drude weight, respectively.
The ratio between the two conductivities, which determines the current responsivity, the figure of merit for rectification from the NLH effect~\cite{2020IsobeRectification}, has been considered a purely intrinsic quantity determined only by the electronic structure~\cite{2021ZhangPNAS}.

In real systems, however, scattering does not simply relax the driven carriers to the equilibrium as assumed in the CRTA; its microscopic detail determines the non-equilibrium distribution of the carriers.
Consequently, CRTA breaks the conservation of total charge~\cite{1970Mermin} and completely misses the emergent quasi-conservation of quantities such as total momentum~\cite{MahanBook,2015Cepellotti,2022Mu}, which manifests as enhanced lifetimes of such quantities.
For example, CRTA does not distinguish forward and backward scatterings, even though only the latter dissipate electric current and contribute to the momentum lifetime~\cite{MahanBook}.
The disparate timescales of the dynamics can be captured only by solving the full Boltzmann transport equation (BTE) which takes into account vertex corrections to the conductivity~\cite{2019KimVertex}.
Many studies have reported significant vertex corrections to linear transport properties, including the linear magnetoresistance~\cite{2009YybornyMR} and the linear anomalous Hall effect~\cite{2017XiaoAHE} of the Rashba Hamiltonian, and the linear mobility~\cite{2016ZhouMobility, 2018PonceMobility, 2020PonceReview, 2021PonceMobility, 2021DesaiMobility, 2022ClaesMobility} and spin lifetimes~\cite{2022ParkSpinLifetime,2022ParkSpinLifetime2} of real materials under electron-phonon scattering.
However, it is still unclear to what extent and through which long-lived quantity vertex correction affects the intrinsic NLH effect.

To gain a physical understanding of the nature of the vertex correction, one can use the relaxon method~\cite{1970Hardy,2016CepellottiRelaxon}.
Relaxons are normal modes of the dynamics described by BTE with well-defined lifetimes.
The relaxon method has been applied to study the quasi-conservation of the total momentum of phonons~\cite{2016CepellottiRelaxon,2020SimoncelliRelaxon}.
However, for electronic transport, quantities other than the total momentum may become long-lived.
Whether the relaxon method could be utilized for electronic systems to investigate long-lived quantities remains an open question.
Although the electron relaxon method was used to compute the linear conductivity~\cite{2022CepellottiPhoebe}, it has not been used for the NLH conductivity, and, much more importantly, neither the analysis of each relaxon mode and its the contribution to the linear or NLH conductivity nor the physical intuition we obtain therefrom has been reported.

In this paper, we report a large contribution to the intrinsic NLH effect that is not captured by the Berry curvature dipole picture.
By exactly solving the linearized BTE for hole-doped GeTe, we find that the current responsivity is up to 60\% larger than that predicted by the CRTA, and exhibits a sharp frequency dependence.
To uncover the origin of these results, we develop a theory of electronic relaxons and find a novel structure in their eigenspectrum and eigenstates.
Based on the relaxon analysis, we attribute the enhancement and frequency dependence to a long-lived difference in the carrier population between the valleys or the minivalleys [Fig.~\ref{fig:schematic}(b)].

\begin{figure}[tb]
\centering
\includegraphics[width=1.0\columnwidth]{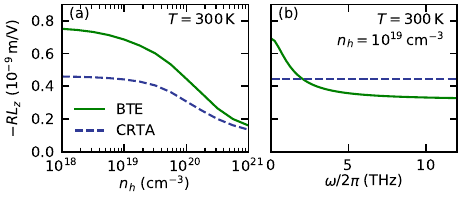}
\caption{
(a) Hole-density dependence, and (b) frequency dependence of the current responsivity multiplied by the sample length $L_z$ for GeTe, which equals the ratio of the NLH conductivity to the linear conductivity [\Eq{eq:responsivity}].
All the results shown in this and the following figures were obtained at hole density $n_h=10^{19}\,\mathrm{cm}^{-3}$ and temperature $T$=300~K unless otherwise noted.
}
\label{fig:conductivity}
\end{figure}

The BTE describes the time evolution of the electron occupation at time $t$.
We write $f_i(t) = f_i^{(0)} + \df_i(t)$, where $f_i^{(0)}$ is the equilibrium Fermi-Dirac occupation in the absence of the external electric field and $i$ the electron eigenstate index specifying both the band and the momentum.
In the linear response regime for a uniform, monochromatic electric field $\mb{E}(t) = \mathrm{Re}\, \mb{E}\, e^{i\omega t}$, the BTE reads as~\cite{MahanBook}
\begin{equation} \label{eq:bte}
\begin{gathered}
    \df_i(t) = \mathrm{Re} \sum_{b=x,y,z} E^b \,\df^b_i (\omega)\, e^{i\omega t},
    \\
    i\omega\, \df^b_i(\omega) = e\, v^b_i\, f^{(0)\prime}_i - \sum_{j} S_{ij}\, \df^b_j(\omega)\,.
\end{gathered}
\end{equation}
Here, $-e$ is the electron charge, $v^b_i$ the band velocity, $f^{(0)\prime}_i$ the energy derivative of the Fermi-Dirac distribution, and $S$ the scattering matrix [\myeqref{eq:scattering}].
Setting $\omega=0$ gives the dc BTE.
The diagonal part of the scattering matrix is the inverse of the single-particle lifetime: $S_{ii} = \tau_i^{-1}$.
In the CRTA, $S$ is approximated to be proportional to the identity matrix, $S^{\rm CRTA}_{ij} = \tau^{-1}\, \delta_{ij}$, where the inverse lifetime $\tau^{-1}$ is a free parameter.
In this work, we focus on the electron-phonon scattering~\cite{ZimanBook,2020PonceReview}.

From the solution of the BTE, one can compute the linear and nonlinear conductivities as
\begin{equation} \label{eq:sigma}
\begin{gathered}
    \sigma^{ab}_{\rm L}(\omega) = -\frac{e}{V_{\rm cell}\, N_k} \sum_i v^a_i\, \df^b_i(\omega), \\
    \sigma^{a;bc}_{\rm NLH}(\omega) = \frac{e^2}{2\,\hbar\, V_{\rm cell}\, N_k} \sum_{d} \epsilon_{adc} \sum_i \Omega^d_i \, \df^b_i(\omega),
\end{gathered}
\end{equation}
where $a$, $b$, $c$, and $d$ are Cartesian indices, $\Omega^d_i$ the band Berry curvature, $\epsilon_{acd}$ the Levi-Civita symbol, $V_{\rm cell}$ the volume of the unit cell, and $N_k$ the number of $k$ points in the full $k$-point grid.
(See Sec.~\ref{sec:SM_bte_and_rta} and Sec.~\ref{sec:comp_details} of the SM for details of the BTE formalism and computational details, respectively~\cite{\citeSupp}.)

We study the linear and nonlinear conductivities of $p$-doped $\alpha$-GeTe, a ferroelectric semiconductor with a large Rashba-type band splitting~\cite{2013DiSanteGeTe}.
Below the transition temperature of 720~K, the Te atom shifts along the [111] direction, which we choose to be the $z$ axis, creating a nonzero polarization.
The broken inversion symmetry allows the NLH response, and GeTe has been proposed as a candidate material for terahertz photodetection using the NLH effect~\cite{2021ZhangPNAS}.
GeTe is typically heavily $p$ doped due to the thermodynamically favorable Ge vacancies~\cite{2005EdwardsGeTe}.
Hole concentrations as low as $5\times10^{19}\,\mathrm{cm}^{-3}$ or below have been realized by additional doping~\cite{2017LiGeTeDoping,2019BuGeTeDoping,2019DongGeTeDoping}.
Doping was treated using the rigid-band approximation, and the free-carrier screening to electron-phonon coupling was included~\cite{2017Verdi}.

We consider the in-plane linear conductivity $\sigma_{\rm L}^{xx}$ and the $\sigma_{\rm NLH}^{z;xx}$ component of the NLH conductivity, unless otherwise stated (see Fig.~\ref{fig:structure} for the crystal structure and the field and current directions).
The linear conductivity is isotropic along the in-plane directions: $\sigma_{\rm L}^{xx} = \sigma_{\rm L}^{yy} \neq \sigma_{\rm L}^{zz}$.
All nonzero components of the NLH conductivity are related by symmetry:
$\sigma_{\rm NLH}^{z;xx} = \sigma_{\rm NLH}^{z;yy} = -\sigma_{\rm NLH}^{x;xz} = -\sigma_{\rm NLH}^{y;yz}$.
We focus on the current responsivity, the figure of merit for terahertz rectification~\cite{2020IsobeRectification}.
For a sample with size $L_x \times L_y \times L_z$, the current responsivitiy $R$ is~\cite{2020IsobeRectification}
\begin{equation} \label{eq:responsivity}
    R = \frac{J^z_{\rm NLH} \, L_x\, L_y}{J^x_{\rm L} \, E^x \, L_x L_y L_z}
    = \frac{1}{L_z} \frac{\sigma_{\rm NLH}^{z;xx}}{\sigma_{\rm L}^{xx}}\,.
\end{equation}

Figure~\ref{fig:conductivity}(a) shows that for a wide range of carrier densities, the current responsivity calculated with BTE is considerably larger than that obtained in the CRTA.
Figure~\ref{fig:conductivity}(b) shows that the ac responsivity in the terahertz regime is sharply peaked at zero frequency and rapidly drops at higher frequencies, while the responsivity in the CRTA is frequency independent.
In Sec.~\ref{sec:supp_temperature} of the SM~\cite{\citeSupp}, we show that this feature is present for a wide range of temperatures.
These findings demonstrate a substantial vertex correction to the NLH effect, and suggest that an important mechanism for the intrinsic NLH effect beyond the conventional Berry curvature dipole picture exists.

\begin{figure}[tb]
\centering
\includegraphics[width=1.0\columnwidth]{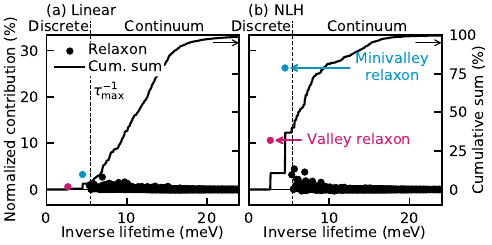}
\caption{
Relaxon decomposition of the (a) linear conductivity and (b) NLH conductivity of GeTe.
We only show individual contributions whose absolute value is greater than $10^{-3}\,$\%.
}
\label{fig:relaxon_decomp}
\end{figure}

\begin{figure}[htb]
\centering
\includegraphics[width=1.0\columnwidth]{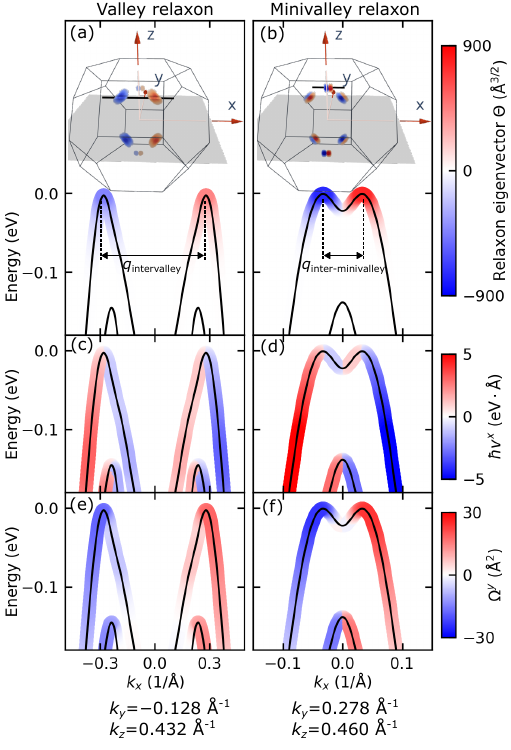}
\caption{
Relaxon eigenvectors, velocity, and Berry curvature of GeTe.
(a, b) Relaxon eigenvector [$\Theta_{ip}$ in \Eq{eq:stilde_diag}] , (c, d) electron band velocity, and (e, f) electron Berry curvature, along the solid black line in the inset of (a, b).
Inset: momentum-space distribution of the relaxon at the highest-energy valence band.
The grey shading indicates the $xy$ plane.
}
\label{fig:relaxon_dist}
\end{figure}

To uncover the mechanism underlying the large increase and frequency dependence of the NLH conductivity, we generalize the relaxon decomposition, originally developed for phonon transport~\cite{1970Hardy,2016CepellottiRelaxon}, to the case of electrons.
Relaxons are normal modes of the BTE, which can be obtained via the eigenmodes of the scattering matrix $S$.
Since $S$ is not symmetric, one first defines the symmetrized scattering matrix $\widetilde{S}$ by applying a similarity transformation~\cite{2022CepellottiPhoebe}
\begin{equation} \label{eq:stilde_def}
    \widetilde{S}_{ij} = S_{ij} \sqrt{\frac{f^{(0)}_j \left(1 - f^{(0)}_j\right)}{f^{(0)}_i \left(1 - f^{(0)}_i\right)}}\,.
\end{equation}
This matrix is real-valued, symmetric, and positive semi-definite, and has the same set of eigenvalues as $S$.
(We note that other types of symmetric scattering matrices, which are \textit{not} related to $S$ by similarity transformations, have also been used~\cite{Fiorentini2016,Macheda2018,Macheda2020}.)
We diagonalize $\widetilde{S}$ as
\begin{equation} \label{eq:stilde_diag}
    \sum_j \widetilde{S}_{ij} \, \Theta_{jp} = \Gamma_p \, \Theta_{ip},
\end{equation}
with the normalization condition $(\Theta^\intercal\, \Theta)_{pp'} = V_{\rm cell}\, N_k\, \delta_{pp'}$.
Each column $\Theta_{ip}$ describes a relaxon mode whose relaxation rate or inverse lifetime is $\Gamma_p$.
Each eigenvalue $\Gamma_p^{-1}$ is a non-negative, physical lifetime of a collective mode.
One of the relaxons has a zero eigenvalue and corresponds to the change of the chemical potential [\Eq{eq:Theta_0}]; the remaining relaxons have positive eigenvalues.

By writing the stationary-state electron distribution in the relaxon basis, one can decompose the conductivities into the contributions of individual relaxons:
\begin{equation} \label{eq:sigma_relaxon}
\begin{gathered}
    \Re \sigma^{ab}_{\rm L}(\omega)
    = e^2\, \kappa\, \sum_p \widetilde{v}^a_p \, \widetilde{v}^b_p \, \frac{\Gamma_p}{\Gamma_p^2 + \omega^2}, \\
    \Re \sigma^{a;bc}_{\rm NLH}(\omega)
    = -\frac{e^3\, \kappa}{2\,\hbar} \sum_{d} \epsilon_{adc}\, \sum_p \widetilde{\Omega}^d_p \, \widetilde{v}^b_p \, \frac{\Gamma_p}{\Gamma_p^2 + \omega^2}.
\end{gathered}
\end{equation}
Here, $\kappa$ is the charge compressibility, and $\widetilde{v}^a_p$ and $\widetilde{\Omega}^a_p$ are the effective velocity and Berry curvature of the relaxons, respectively, which are the averages of the quantities over the electron eigenstates weighted by the relaxon eigenvector [Eqs.~(\ref{eq:relaxon_v}, \ref{eq:relaxon_Omega})].
The charge compressibility can be understood as the charge carried by a single relaxon, which is a collective excitation spanning the whole Fermi surface.
For heat transport by phonon relaxons, the charge compressibility is replaced with the heat capacity~\cite{2016CepellottiRelaxon}.

Figure~\ref{fig:relaxon_decomp} shows the contribution of each relaxon to the linear and NLH conductivities.
The relaxon spectrum has two parts, discrete levels and a continuum, separated by the inverse of $\tau_{\rm max} = \max_i \tau_i$, the longest single-particle lifetime.
Relaxons with lifetimes longer than $\tau_{\rm max}$ ($\Gamma_p^{-1} > \tau_{\rm max}$) display a discrete spectrum, while those with lifetimes shorter than $\tau_{\rm max}$ ($\Gamma_p^{-1} < \tau_{\rm max}$) form a continuum.
This spectral structure of relaxons resembles that of optical excitations of a hydrogen atom or of semiconductors (discrete excitons and a continuum)~\cite{YuCardonaBook}, although the former corresponds to inverse lifetimes while the latter correspond to energies.
Also, we find that the effective velocity and Berry curvature scale as $\mcO(1)$ and $\mcO(1/\sqrt{N_k})$ for the discrete and continuum relaxons, respectively.
This scaling, combined with the $\mcO(1)$ ($\mcO(N_k)$) scaling of the number of discrete (continuum) relaxons and the $\mcO(1)$ scaling of the relaxon lifetimes, makes the sum over relaxons in \Eq{eq:sigma_relaxon} convergent.
As one increases the density of the $k$-point grid, the cumulative contribution remains discontinuous in the discrete regime and converges to a smooth curve in the continuum regime.
See Sec.~\ref{sec:SM_spectra} and Fig.~\ref{fig:relaxon_nkdep} of the SM for details~\cite{\citeSupp}.

In Fig.~\ref{fig:relaxon_decomp}, among the contributions of more than 10,000 relaxons to the conductivities, those of the two discrete, long-lived relaxons labelled ``valley relaxon'' and ``minivalley relaxon'' stand out.
Remarkably, although these two relaxons contribute only 4\% to the linear conductivity, they account for 37\% of the NLH conductivity.
We have named these relaxons based on their eigenstates shown in Fig.~\ref{fig:relaxon_dist}.
The valley relaxon has a strong valley-polarizing character, describing the transfer of hole carriers from one valley to another valley.
Similarly, the minivalley relaxon has a minivalley-polarizing character, in which ``minivalley'' refers to each of the two peaks of the Mexican-hat-like band structure within a single valley [Fig.~\ref{fig:relaxon_dist}(b)].

The eigenvectors of the valley and minivalley relaxons are peaked around the valence band maxima, with the same sign within a (mini)valley and the opposite signs between different (mini)valleys.
Owing to the quadratic dispersion, the band velocities around the valence band maxima are small and approximately antisymmetric in wavevector space within a (mini)valley (Fig.~\ref{fig:relaxon_dist}(c,d)).
Hence, the effective velocity of the (mini)valley relaxons is low.
In contrast, the Berry curvature is large near the valence band maxima and has the same parity in momentum space as the relaxon eigenvector (Fig.~\ref{fig:relaxon_dist}(e,f)).
Thus, both the valley and minivalley relaxons strongly couple to the Berry curvature, resulting in their large contribution to the NLH effect.
This valley-polarization induced NLH effect will be present in all inversion-asymmetric materials with multiple valleys, such as the two-dimensional transition metal dichalcogenides~\cite{2014MakScienceValley}.

The frequency dependence of the current responsivity [Fig.~\ref{fig:conductivity}(b)] can be understood from the long lifetimes of the valley-polarizing relaxons, which originate from the weak intervalley scattering.
Each relaxon gives a Drude-like contribution to the ac conductivity: a Lorentzian with a width $\Gamma_p$.
Since the long-lived valley and minivalley relaxons contribute much more to the NLH conductivity than to the linear conductivity, the ac spectrum of the former is more sharply peaked at $\omega=0$ than the latter.
Hence, the current responsivity rapidly decays at higher frequencies as the contribution from the valley-polarizing relaxons decays faster than the remainder.
Terahertz measurements could experimentally confirm this prediction.

It may be surprising that the valley and minivalley relaxons are highly populated in the steady state under a uniform electric field, since a pure valley polarization by itself does not generate electric currents or couple to the electric field.
In fact, the valley polarization vanishes in the CRTA because $\df$ becomes a $k$-derivative of the equilibrium occupation, whose integral over a valley is zero.
The emergence of valley polarization is a result of the interplay of multiple intervalley scattering channels, as shown in a study of twisted bilayer graphene using a model with two circular Fermi surfaces and scattering matrix elements that vary sinusoidally with the electron wavevector direction~\cite{2021YingValleyPol}.

\begin{figure}[tb]
\centering
\includegraphics[width=0.78\columnwidth]{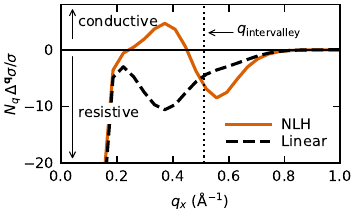}
\caption{
Scattering diagnostics for GeTe along $\mb{q}=(q_x,0,0)$.
$q_{\rm intervalley}$ is the intervalley distance (Fig.~\ref{fig:relaxon_dist}(a)).
}
\label{fig:diagnostics_1d_zoom}
\end{figure}

To study which phonon modes contribute to the valley polarization, we propose and apply the ``scattering diagnostics'' method.
We solve a modified BTE in which phonons with momentum $\mb{q}$ are excluded in the construction of the scattering matrix and see the change in conductivity, $\Delta^{\mb{q}} \sigma$, as we add back the missing scattering, i.\,e.\,, recover the conductivity $\sigma$ (see Sec.~\ref{sec:SM_diagnostics} of the SM for details~\cite{\citeSupp}).
A negative value of $\Delta^\mb{q} \sigma / \sigma$ indicates that the scattering tends to suppress the conductivity, while a positive value indicates that the scattering enhances the conductivity, which would be an uncommon situation.

Figure~\ref{fig:diagnostics_1d_zoom} shows the scattering diagnostics for $q$ points along the field direction.
For the linear conductivity, $\Delta^{\mb{q}} \sigma$ is always negative, indicating that all scattering is resistive.
In contrast, for the NLH effect, phonons with momentum close to but below the intervalley distance $q_{\rm intervalley}$ increase the NLH conductivity.
This behavior can be understood in terms of a four-state model (Fig.~\ref{fig:SM_scat_model}), which shows that large valley polarization can occur when the intervalley scattering rate is highly momentum-dependent.
For GeTe, the intervalley scatterings with momentum smaller than the intervalley distance, $q \lesssim q_{\rm intervalley}$, are stronger than those with momentum larger than the intervalley distance, $q \gtrsim q_{\rm intervalley}$.
Scattering by phonons with $q \lesssim q_{\rm intervalley}$ ($q \gtrsim q_{\rm intervalley}$) strengthens (weakens) this momentum dependence, thus enhancing (suppressing) the valley polarization and the NLH effect.

A similar analysis could be performed for the minivalley relaxons, based on the momentum dependence of the inter-minivalley scattering.
The presence of minivalleys is a result of the Mexican-hat-like band structure of GeTe.
We expect that the minivalley relaxon will also appear in other materials with a similar dispersion, such as the two-dimensional chalcogenides GaS, GaSe, and InSe~\cite{2014Rybkovskiy,2018Kibirev,2019Li}.

Scattering-induced valley polarization is a general phenomenon that occurs in many polar and nonpolar semiconductors that have multiple valleys at momenta which are not time-reversal invariant (see Sec.~\ref{sec:SM_valley_pol} of the SM for precise symmetry constraints~\cite{\citeSupp}).
The valley polarizability, valley polarization per current, of four common semiconductors, diamond, silicon, GaP, and AlSb, are all of similar order of magnitude to that of GeTe
(Fig.~\ref{fig:valley_pol}).

Recently, Ref.~\cite{2021DuVertex} reported that the vertex correction to the intrinsic NLH conductivity is negligible for the toy model of two-dimensional Dirac electrons with a single valley and an isotropic scattering mechanism.
Our first-principles study suggests that real materials would instead exhibit a substantial vertex correction when the detailed electronic structure with multiple (mini)valleys and momentum-dependent electron-phonon scattering is taken into account.
Finding materials with multiple valleys and momentum-dependent scattering represents a new design principle for maximizing NLH efficiency.

The scattering-induced valley polarization affects not only the NLH effect, but also all observables that couple to the valley degree of freedom.
Such observables include the spin polarization~\cite{2012XiaoPRLValley}, orbital magnetization~\cite{2007XiaoPRLValley}, and circular dichroism~\cite{2008WangPRBValley}.
The interplay of the valley-polarizing relaxons with these degrees of freedom is an interesting direction for valleytronics research~\cite{2016SchaibleyValleytronics}.
For example, circularly polarized luminescence~\cite{2012MakValley,2012ZengValley} could be used to directly measure the valley-polarizing relaxons.

Moreover, our work shows that solving the full BTE and performing the relaxon analysis is a computationally tractable way to reveal emergent long-lived quantities in electronic dynamics and study how they affect the nonlinear responses.
By combining the relaxon analysis and the theories of nonlinear transport including band-geometric effects~\cite{2019MatsyshynPRL,2019ParkerDiagram,2020Holder,2021DuVertex,2022Ahn} using \emph{ab initio} scattering matrices~\cite{2016PonceEPW,2021ZhouPerturbo}, one can reveal novel response properties of real materials beyond the RTA, as demonstrated in this work for the NLH effect.

\begin{acknowledgments}
This work was supported by the Creative-Pioneering Research Program through Seoul National University, Korean NRF No-2023R1A2C1007297, and the Institute for Basic Science (No. IBSR009-D1).
Computational resources have been provided by KISTI (KSC-2021-CRE-0573).
\end{acknowledgments}

\FloatBarrier 

\makeatletter\@input{xy.tex}\makeatother
\bibliography{main}

\end{document}


\title{Supplementary information: Nonlinear Hall effect from long-lived valley-polarizing relaxons}

\author{Jae-Mo Lihm}
\email{jaemo.lihm@gmail.com}
\author{Cheol-Hwan Park}
\email{cheolhwan@snu.ac.kr}
\affiliation{Department of Physics and Astronomy, Seoul National University, Seoul 08826, Korea,}
\affiliation{Center for Correlated Electron Systems, Institute for Basic Science, Seoul 08826, Korea,}
\affiliation{Center for Theoretical Physics, Seoul National University, Seoul 08826, Korea}

\date{\today}

\maketitle



\newpage
\title{Supplementary Information: Nonlinear Hall effect from long-lived, valley-polarizing relaxons}
\maketitle

\appendix

\renewcommand{\thefigure}{S\arabic{figure}}
\setcounter{figure}{0}
\renewcommand{\thetable}{S\arabic{table}}
\setcounter{table}{0}

\section{Boltzmann transport equation and the relaxon decomposition} \label{sec:SM_bte_and_rta}

In the presence of a uniform, time-dependent electric field $\mb{E}(t)$, the BTE in the linear response regime reads~\cite{MahanBook}
\begin{equation} \label{eq:SM_bte}
    \frac{\partial \, \df_i(t)}{\partial t} = e\mb{E}(t) \cdot \mb{v}_i \, f^{(0)\prime}_i - \sum_{j} S_{ij} \,\df_j(t).
\end{equation}
We use the shorthand notation for the electron eigenstate indices $i = (n,\mb{k})$ and $j = (n',\mb{k}')$.
The scattering matrix $S$ is defined as
\begin{equation} \label{eq:scattering}
    S_{ij} = \tau^{-1}_i \delta_{ij} - \tau^{-1}_{i \leftarrow j} (1 - \delta_{ij})
\end{equation}
where $\tau^{-1}_{j \leftarrow i}$ is the rate of the scattering from state $i$ to $j$, and
\begin{equation} \label{eq:tau_inv_def}
    \tau^{-1}_i = \sum_{j \neq i} \tau^{-1}_{j \leftarrow i}
\end{equation}
is the quasiparticle lifetime of state $i$.
The first term on the right-hand side of \myeqref{eq:scattering} describes the ``scattering-out'' process in which the electron at state $i$ decays to another state, while the second term describes the ``scattering-in" process in which an electron from another state scatter into state $i$.

The partial decay rate for electron-phonon scattering is~\cite{ZimanBook,2020PonceReview}
\begin{align} \label{eq:scattering_mel}
    \tau^{-1}_{i \leftarrow j} = \frac{2\pi}{\hbar N_k} &\sum_\nu \abs{g_{ji\nu}}^2
    \Big[ \delta(\veps_i - \veps_j - \hbar\omega_\qnu) (n^{(0)}_\qnu + f^{(0)}_i) \nnnl
        + & \delta(\veps_i - \veps_j + \hbar\omega_\qnu) (n^{(0)}_\qnu + 1 - f^{(0)}_i)\Big],
\end{align}
where 
$\varepsilon_i$ is the electron energy,
$\mb{q}=\mb{k}'-\mb{k}$ the momentum transferred to the phonon,
$\omega_\qnu$ the phonon energy, $n^{(0)}_\qnu$ the phonon occupation number, and
\begin{equation}
    g_{ji\nu} \equiv g_{n'n\nu}(\mb{k},\mb{q}) = \mel{n'\mb{k}+\mb{q}}{\partial_\qnu V}{n\mb{k}}
\end{equation}
the electron-phonon matrix element.
Here, we have defined $\ket{n\mb{k}}$ the electron eigenstate with momentum $\mb{k}$ and band $n$, and
\begin{equation}
    \partial_\qnu V = \sum_{\ell, \kappa, \alpha} \sqrt{\frac{\hbar}{2M_\kappa \omega_\qnu}} e^{i\mb{q}\cdot\mb{R}_\ell} u_{\kappa\alpha,\nu}(\mb{q}) \frac{\partial V}{\partial R_{\ell\kappa\alpha}}
\end{equation}
the phonon perturbation potential where $R_{\ell\kappa\alpha}$ is the $\alpha$-th Cartesian component of the position of the atom $\kappa$ in the $\ell$-th unit cell, $M_\kappa$ the mass of the atom, and $u_{\kappa\alpha,\nu}(\mb{q})$ the phonon eigendisplacement.
We assume that the phonon occupation is fixed to its value at thermal equilibrium, i.\,e.\,, we ignore the phonon drag effect~\cite{2015ZhouDrag,Fiorentini2016,2020ProtikDrag}.

The ordinary linear current for a given carrier distribution $\df_i$ is
\begin{equation} \label{eq:bte_J_linear}
    \mb{J}_{\rm L}(t) = -\frac{e}{V_{\rm cell} \, N_k} \sum_i \mb{v}_i \, \df_i(t).
\end{equation}
The expression for the NLH current, which leads to the NLH effect, follows by replacing $\mb{v}_i$ with the anomalous velocity $\mb{v}_{i,{\rm anom.}}$~\cite{2010XiaoRMP}:
\begin{equation} \label{eq:bte_J_NLH}
    \mb{J}_{\rm NLH}(t) = -\frac{e}{V_{\rm cell} \, N_k} \sum_i \mb{v}_{i,{\rm anom.}} \, \df_i(t)\,,
\end{equation}
where
\begin{equation} \label{eq:v_anom}
    \mb{v}_{i,{\rm anom.}}=\frac{e}{\hbar} \,\mb{E} \times \mb{\Omega}_i\,.
\end{equation}
Here, the Berry curvature $\mb{\Omega}_i$ is defined as
\begin{equation} \label{eq:berry_curvature_def}
    \mb{\Omega}_\nk = -\Im \sum_m \mb{A}_{nm\mb{k}} \times \mb{A}_{mn\mb{k}}
\end{equation}
where
\begin{equation}
    \mb{A}_{mn\mb{k}}= i\braket{u_\mk}{\frac{\partial u_\nk}{\partial \mb{k}}}
\end{equation}
is the Berry connection with $\ket{u_\mk}$ the periodic part of the Bloch eigenstate.
Note that the $\mathcal{O}(E^2)$ term of the occupation function is even in $\mb{k}$ and thus does not contribute to the conductivity of a time-reversal symmetric material whose band velocity and Berry curvature are odd in $\mb{k}$~\cite{2015SodemannPRL}.

Let us consider a monochromatic electric field ($\omega = 0$ yields the DC limit)
\begin{equation}
    \mb{E}(t) = \Re \mb{E}_\omega \, e^{i\omega t},
\end{equation}
where the linear response occupation can be written as
\begin{equation}
    \df_i(t) = \Re \sum_b E_\omega^b \, \df^b_i(\omega) \, e^{i\omega t}.
\end{equation}
Then, the BTE [\myeqref{eq:SM_bte}] becomes
\begin{equation} \label{eq:ac_bte}
    i\omega \df^b_i(\omega) = e \, v^b_i \, f^{(0)\prime}_i - \sum_{j} S_{ij} \, \df^b_j(\omega).
\end{equation}
Given the occupation, the linear current [\Eq{eq:bte_J_linear}] reads
\begin{align}
    J^a_{\rm L}(t)
    &= -\frac{e}{V_{\rm cell} \, N_k} \sum_i v^a_i\, \Re \sum_b E_\omega^b \, \df^b_i(\omega) \, e^{i\omega t} \nnnl
    &= \Re \sum_b \sigma^{ab}_{\rm L} \, E^b_\omega \, e^{i\omega t},
\end{align}
with the linear conductivity
\begin{equation} \label{eq:ac_sigma_l}
    \sigma^{ab}_{\rm L}(\omega) = -\frac{e}{V_{\rm cell} \, N_k} \sum_i v^a_i \, \df^b_i(\omega).
\end{equation}

For the NLH current, we modify \Eqs{eq:bte_J_NLH} and \eqref{eq:berry_curvature_def} to use the finite-frequency generalization of the Berry curvature~\cite{2018TsirkinGyrotropic}:
\begin{equation} \label{eq:ac_berry}
    \mb{\Omega}_\nk(\omega) = - \Im \sum_m \frac{(\varepsilon_\mk - \varepsilon_\nk)^2}{(\varepsilon_\mk - \varepsilon_\nk)^2 - \hbar^2 \omega^2}\, \mb{A}_{nm\mb{k}} \times \mb{A}_{mn\mb{k}}.
\end{equation}
Then, the NLH current is
\begin{align}
    &\quad J^a_{\rm NLH}(t) \nnnl
    &= -\frac{e^2}{\hbar\, V_{\rm cell} \, N_k} \sum_i (\mb{E}(t) \times \mb{\Omega}_i(\omega))^a \Re \sum_b E_\omega^b \, \df^b_i(\omega) \, e^{i\omega t} \nnnl
    &= -\frac{e^2}{2\,\hbar\, V_{\rm cell} \, N_k} \sum_i \sum_{c,d} \Big \{ \epsilon_{acd}\, [E^c_\omega \, e^{i\omega t} + (E^c_\omega)^* e^{-i\omega t}] \, \Omega^d_i(\omega) \nnnl
    &\quad\quad \times \Re \sum_b E_\omega^b \, \df^b_i(\omega) \, e^{i\omega t} \Big\} \nnnl
    &= \Re \sum_{b,c} \sigma^{a;bc}_{\rm NLH}(\omega)\,
    \big[ E_\omega^b \, E^c_\omega \, e^{2i\omega t} + E_\omega^b \,(E^c_\omega)^* \big]
\end{align}
with the AC NLH conductivity
\begin{equation} \label{eq:ac_sigma_nl}
    \sigma^{a;bc}_{\rm NLH}(\omega) = \frac{e^2}{2\,\hbar\, V_{\rm cell} \, N_k} \sum_{d} \epsilon_{adc} \sum_i \Omega^d_i(\omega) \, \df^b_i(\omega).
\end{equation}
Note that the anomalous current has two terms, the second-harmonic term with frequency $2\omega$ and the static rectification term.
In the BTE, the two terms share the same conductivity tensor.

Now, we study the solution of the BTE in the relaxon basis.
By construction, $x_i = 1$ is a left eigenvector of the scattering matrix with eigenvalue zero:
\begin{equation} \label{eq:charge_cons}
    \sum_i S_{ij} = \tau_j^{-1} - \sum_{i \neq j} \tau^{-1}_{i \leftarrow j} = 0.
\end{equation}
This property easily follows from \Eqs{eq:scattering} and \eqref{eq:tau_inv_def}.
For the symmetrized scattering matrix $\widetilde{S}$ [\Eq{eq:stilde_def}], the corresponding null eigenvector is
\begin{equation} \label{eq:Theta_0}
    \Theta_{i0} = \frac{1}{\sqrt{k_{\rm B} \, T \, \kappa}} \sqrt{f^{(0)}_i \left( 1 - f^{(0)}_i \right)}
    = \sqrt{-\frac{1}{\kappa} f^{(0)'}_i}\,.
\end{equation}
where
\begin{align} \label{eq:compressibility}
    \kappa = \frac{d n}{d\mu}
    &= \frac{1}{V_{\rm cell}\, N_k} \sum_i \frac{d f^{(0)}_i}{d\mu} \nnnl
    &= \frac{1}{k_{\rm B} \, T} \frac{1}{V_{\rm cell}\, N_k} \sum_i f^{(0)}_i \left( 1 - f^{(0)}_i \right)
\end{align}
is the charge compressibility.
This eigenvector, which is one of the relaxons, satisfies the normalization condition
\begin{equation}
    \frac{1}{V_{\rm cell}\, N_k} \sum_i \Theta_{i0}^2
    = \frac{1}{k_{\rm B} \, T \, \kappa} \frac{1}{V_{\rm cell}\, N_k} 
 \sum_i f^{(0)}_i \left( 1 - f^{(0)}_i \right)
    = 1.
\end{equation}
This relaxon describes the change in the electron occupation due to the increase in the chemical potential, $\df_i \propto df^{(0)}_i / d\mu$.
Such a perturbation to the occupation cannot be dissipated because of charge conservation.
Hence, the relaxon has a zero scattering rate.

We now define the effective velocity~\cite{1970Hardy,2016CepellottiRelaxon} and the Berry curvature of the relaxons, which is the sum of the band velocity and the Berry curvature weighted by the relaxon eigenvector and $\Theta_{i0}$:
\begin{align}
    \label{eq:relaxon_v}
    \widetilde{\mb{v}}_p
    &\equiv \frac{1}{V_{\rm cell} \, N_k} \sum_i \mb{v}_i  \,\Theta_{i0}\, \Theta_{ip} \nnnl
    &= \frac{1}{V_{\rm cell} \, N_k} \frac{1}{\sqrt{k_{\rm B} \, T \, \kappa}} \sum_i \mb{v}_i\, \sqrt{f^{(0)}_i \left( 1 - f^{(0)}_i \right)}\, \Theta_{ip}\nnnl
    &=\frac{\sum_i \mb{v}_i \,\sqrt{f^{(0)}_i \left( 1 - f^{(0)}_i \right)}\, \left(\Theta_{ip}/\sqrt{V_{\rm cell} \, N_k}\right)}{\sqrt{\sum_i f^{(0)}_i \left( 1 - f^{(0)}_i \right)}}\,,
    \\
    \label{eq:relaxon_Omega}
    \widetilde{\mb{\Omega}}_p
    &\equiv \frac{1}{V_{\rm cell} \, N_k} \sum_i \mb{\Omega}_i\, \Theta_{i0}\, \Theta_{ip} \nnnl
    &= \frac{1}{V_{\rm cell} \, N_k} \frac{1}{\sqrt{k_{\rm B} \, T \, \kappa}} \sum_i \mb{\Omega}_i \sqrt{f^{(0)}_i \left( 1 - f^{(0)}_i \right)} \Theta_{ip}\nnnl
    &=\frac{\sum_i \mb{\Omega}_i \,\sqrt{f^{(0)}_i \left( 1 - f^{(0)}_i \right)}\, \left(\Theta_{ip}/\sqrt{V_{\rm cell} \, N_k}\right)}{\sqrt{\sum_i f^{(0)}_i \left( 1 - f^{(0)}_i \right)}}\,.
\end{align}
The last lines of \Eq{eq:relaxon_v} and \Eq{eq:relaxon_Omega}, which we obtained using \Eq{eq:compressibility}, clearly show that the effective velocity and the Berry curvature of the relaxons are the weighted average of those quantities over the thermally activated single-particle Bloch states near the Fermi level.
Also, from the normalization convention of relaxons,
these effective quantities have the same units as those of the single-particle eigenstates.
Our normalization convention is consistent with Ref.~\cite{2016CepellottiRelaxon} if the charge compressibility is replaced by the heat capacity.

Now, we rewrite the conductivities in the relaxon basis.
We begin by writing the electron distribution in the relaxon basis:
\begin{equation} \label{eq:ac_bte_f_relaxon}
    \df^b_i(\omega) = \sum_p \sqrt{f^{(0)}_i \left(1 - f^{(0)}_i\right)} \, \Theta_{ip} \, \dftilde^b_p(\omega).
\end{equation}
Transforming the BTE [\myeqref{eq:ac_bte}] to the relaxon basis yields
\begin{equation} \label{eq:ac_bte_relaxon}
    i\omega \,\dftilde^b_p(\omega) = -e \,\sqrt{\frac{\kappa}{k_{\rm B} \, T}} \widetilde{v}_p^b - \Gamma_p \, \dftilde^b_p(\omega).
\end{equation}
By solving this equation, we find the stationary-state relaxon occupation
\begin{equation} \label{eq:ac_relaxon_occupation}
    \dftilde^b_p(\omega) = -e\, \sqrt{\frac{\kappa}{k_{\rm B} \, T}}\, \frac{\widetilde{v}_p^b}{i\omega + \Gamma_p}.
\end{equation}
Transforming back to the electron eigenstate basis, we find
\begin{align} \label{eq:ac_bte_f_relaxon_solution}
    \df^b_i(\omega)
    &= -e\, \sqrt{\frac{\kappa}{k_{\rm B} \, T}}\, \sqrt{f^{(0)}_i \left(1 - f^{(0)}_i\right)} \,\sum_p \Theta_{ip}\, \frac{\widetilde{v}_p^b}{(i\omega + \Gamma_p)} \nnnl
    &= -e\, \kappa\, \Theta_{i0}\, \sum_p \Theta_{ip} \frac{\widetilde{v}_p^b}{(i\omega + \Gamma_p)}\,.
\end{align}
By substituting this occupation into Eqs.~(\ref{eq:ac_sigma_l}, \ref{eq:ac_sigma_nl}), we find the AC conductivities in the relaxon basis:
\begin{equation}
\begin{gathered}
    \sigma^{ab}_{\rm L}(\omega)
    = e^2\, \kappa\, \sum_p \widetilde{v}^a_p \, \widetilde{v}^b_p \, \frac{\Gamma_p -i\omega}{\Gamma_p^2 + \omega^2}, \\
    \sigma^{a;bc}_{\rm NLH}(\omega)
    = -\frac{e^3\, \kappa}{2\,\hbar} \sum_{d} \epsilon_{adc}\, \sum_p \widetilde{\Omega}^d_p(\omega) \, \widetilde{v}^b_p \, \frac{\Gamma_p -i\omega}{\Gamma_p^2 + \omega^2}.
\end{gathered}
\end{equation}
Here, $\widetilde{\Omega}^d(\omega)$ is defined analogously to \myeqref{eq:relaxon_Omega}, using the frequency-dependent Berry curvature $\Omega^d_i(\omega)$ [\Eq{eq:ac_berry}] instead of the static one $\Omega^d_i$ [\Eq{eq:berry_curvature_def}].

\begin{figure}[tb]
\centering
\includegraphics[width=0.99\linewidth]{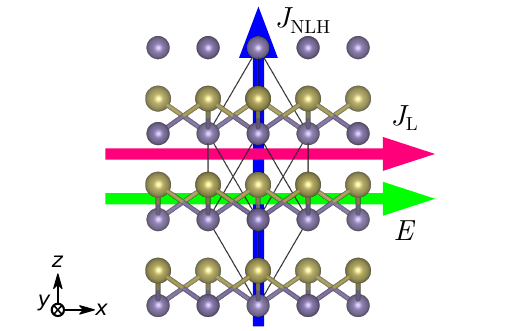}
\caption{
Crystal structure of $\alpha$-GeTe and the field and current directions considered in this work.
}
\label{fig:structure}
\end{figure}

The crystal structure and the field and current directions considered in this work are shown in Fig.~\ref{fig:structure}.

\section{Discrete and continuous eigenvalue spectra of relaxons} \label{sec:SM_spectra}
In this section, we discuss the eigenvalue spectra of the scattering matrix.
In the limit of an infinitely dense $k$-point grid, the eigenvalues of the scattering matrix, i.e., the relaxon eigenvalues, are discrete below $\tau^{-1}_{\rm max} = (\max_i \tau_i)^{-1}$ and continuous above $\tau^{-1}_{\rm max}$.

\begin{figure}[tb]
\centering
\includegraphics[width=0.99\linewidth]{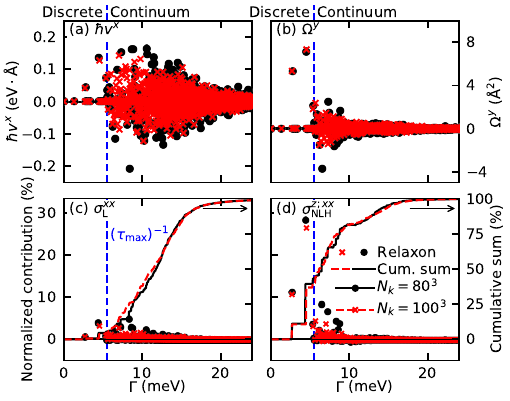}
\caption{
(a) Relaxon effective velocity and (b) relaxon effective Berry curvature obtained for two different $k$-point grids.
(c, d) Relaxon decomposition and its cumulative sum for the (a) linear and (b) NLH conductivity of GeTe.
We plot only the individual contributions whose absolute value is greater than $10^{-3}\,$\%.
}
\label{fig:relaxon_nkdep}
\end{figure}

Figure~\ref{fig:relaxon_nkdep} compares the relaxon velocity [Fig.~\ref{fig:relaxon_nkdep}(a)],
relaxon Berry curvature [Fig.~\ref{fig:relaxon_nkdep}(b)],
and relaxon contribution to the conductivity and its cumulative sum [Fig.~\ref{fig:relaxon_nkdep}(c,d)]
obtained at two different $k$-point grids.
Changing the $k$-point grid from 80$\times$80$\times$80 to 100$\times$100$\times$100 roughly doubles the $k$-point density ($100^3/80^3 = 1.95$).
Figures~\ref{fig:relaxon_nkdep}(a,b) show that the relaxon velocity and Berry curvature are approximately $N_k$-indepdent in the discrete regime, while decreasing with $N_k$ in the continuum regime.
Figures~\ref{fig:relaxon_nkdep}(c,d) show that the individual contributions to the conductivity of the relaxons in the discrete regime ($\Gamma < \tau_\mathrm{max}^{-1}$) remain approximately unchanged, while those in the continuum regime ($\Gamma > \tau_\mathrm{max}^{-1}$) become approximately halved.
The cumulative distribution remains almost invariant as the number of relaxons in the continuum regime is doubled.

\section{Temperature dependence of the NLH effect} \label{sec:supp_temperature}

\begin{figure}[tb]
\centering
\includegraphics[width=0.99\linewidth]{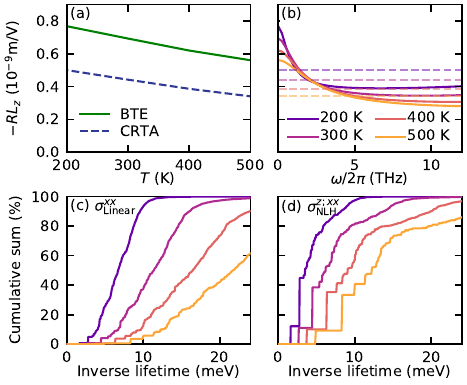}
\caption{
(a) Temperature dependence of the current responsivity of GeTe.
(b) Frequency dependence of the current responsivity at various temperatures.
(c) Cumulative sum of the relaxon-decomposed linear conductivity at various temperatures.
(d) Same as (c), but for the NLH conductivity.
All results are for $n=10^{19}~\mathrm{cm}^{-3}$.
}
\label{fig:temp_dep}
\end{figure}

Figure~\ref{fig:temp_dep} shows the temperature dependence of the NLH effect (current reponsivity is the ratio between the NLH conductivity and the linear conductivity) as well as its relaxon decomposition.
We find that the current responsivitiy decreases with temperature [Fig.~\ref{fig:temp_dep}(a)].
The characteristics of the relaxon decomposition, namely the sharp frequency dependence of the AC current responsivity [Fig.~\ref{fig:temp_dep}(b)],
and the large contribution of the discrete relaxons to the NLH conductivity but not to the linear conductivity [Fig.~\ref{fig:temp_dep}(c,d)] is present at all temperatures.

\section{Scattering diagnostics} \label{sec:SM_diagnostics}
In this section, we develop the ``scattering diagnostics'' method which allows one to investigate how each phonon mode contributes to the conductivity.
The name ``scattering diagnostics'' is inspired by the ``fluctuation diagnostics'' method~\cite{2015GunnarssonFlucDiag} which decomposes self-energy into the contribution of individual fluctuations.
Here, we decompose the scattering matrix into the contributions of individual phonon wavevectors:
\begin{equation}
    S = \sum_{\mb{q}} S^{\mb{q}}.
\end{equation}
To study the effect of a single phonon wavevector $\mb{q}$ on transport, we subtract its contribution from the scattering matrix and solve the modified BTE.
The resulting modified BTE is
\begin{equation} \label{eq:diag_modified_BTE}
    0 = e \, v^b_i \, f^{(0)\prime}_i - \sum_{j} (S_{ij} - S_{ij}^{\mb{q}}) (\df^b_j - \Delta^\mb{q} \df^b_j),
\end{equation}
where $\Delta^\mb{q} \df^b_j$ is the change in the stationary-state electron distribution due to the scattering with wavevector $\mb{q}$.
Solving for $\Delta^\mb{q} \df^b_j$, we find
\begin{equation} \label{eq:diag_modified_BTE_2}
    \sum_j (S_{ij} - S_{ij}^{\mb{q}}) \, \Delta^\mb{q} \df^b_j
    = -\sum_{j} S_{ij}^{\mb{q}} \, \df^b_j.
\end{equation}
Given the solution to this linear equation, we calculate the conductivity using \Eq{eq:sigma}, but with $\Delta^\mb{q} \df^b_j$ instead of $\df^b_j$:
\begin{equation} \label{eq:diag_sigma}
\begin{gathered}
    \Delta^\mb{q} \sigma^{ab}_{\rm L} = -\frac{e}{V_{\rm cell} \, N_k} \sum_i v^a_i \, \Delta^\mb{q} \df^b_i, \\
    \Delta^\mb{q} \sigma^{a;bc}_{\rm NLH} = \frac{e^2}{2\hbar V_{\rm cell} \, N_k} \sum_{d} \epsilon_{adc} \sum_i \Omega^d_i \, \Delta^\mb{q} \df^b_i.
\end{gathered}
\end{equation}
The resulting momentum-resolved conductivity describes how sensitive the conductivity is to the given scattering mechanism.
A negative (positive) value indicates that the scattering tends to suppress (enhance) the conductivity.

If $S^{\mb{q}}$ is much smaller than the total scattering matrix $S$, which is satisfied in the usual setup for first-principles calculations because $N_k \gg 1$, one can approximate \myeqref{eq:diag_modified_BTE_2} by neglecting the $\mathcal{O}((S^{\mb{q}})^2)$ term:
\begin{equation} \label{eq:diag_modified_BTE_3}
    \sum_j S_{ij} \, \Delta^\mb{q} \df^b_j
    \approx -\sum_{j} S_{ij}^{\mb{q}} \, \df^b_j.
\end{equation}
In this approximation, the sum of $\Delta^\mb{q} \df^b_j$ for all momenta $\mb{q}$ satisfies
\begin{equation}
    \sum_j S_{ij} \sum_\mb{q} \Delta^\mb{q} \df^b_j
    \approx -\sum_{j} S_{ij} \, \df^b_j
\end{equation}
and so
\begin{equation}
    \sum_\mb{q} \Delta^\mb{q} \df^b_j = - \df^b_j\,.
\end{equation}
Thus, interestingly, we find
\begin{equation}
    \sum_\mb{q} \Delta^\mb{q} \sigma^{ab}_{\rm L} = -\sigma^{ab}_{\rm L}, \quad
    \sum_\mb{q} \Delta^\mb{q} \sigma^{a;bc}_{\rm NLH} = -\sigma^{a;bc}_{\rm NLH}.
\end{equation}
But, we emphasize that this analysis is based on the subtraction and re-addition of the individual scattering channels to the full scattering matrix.
The effect of all scattering channels cannot be disentangled, and $\sum_\mb{q} \Delta^\mb{q} \sigma$ should not be viewed as the contribution of phonons with wavevector $\mb{q}$ to the scattering.
Instead, it should be understood as a way to quantify how sensitive the conductivity is to the corresponding scattering process.

\begin{figure}[tb]
\centering
\includegraphics[width=1.0\columnwidth]{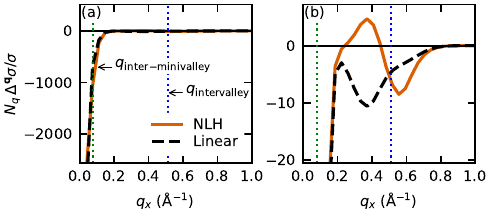}
\caption{
(a) Scattering diagnostics for the linear and NLH conductivities of GeTe for phonons with momentum $\mb{q}=(q_x, 0, 0)$.
(b) Zoom-in of (a).
$N_q$ is the size of the $q$-point grid, which we choose to be identical to $N_k$.
}
\label{fig:diagnostics_1d}
\end{figure}

\begin{figure}[tb]
\centering
\includegraphics[width=1.0\columnwidth]{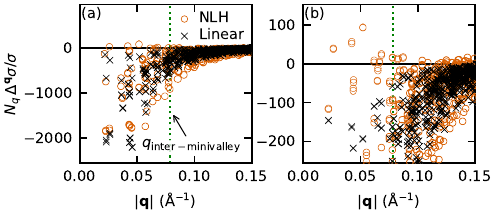}
\caption{
(a) Scattering diagnostics for the linear and NLH conductivities of GeTe for phonons with momentum $q$ near $\Gamma$.
(b) Zoom-in of (a).
}
\label{fig:diagnostics_gamma}
\end{figure}

Figure~\ref{fig:diagnostics_1d} shows the momentum-resolved conductivity along the $q_x$ axis.
From Fig.~\ref{fig:diagnostics_1d}(a), we see that the small-momentum phonon modes suppress the conductivity the most, as in many polar semiconductors~\cite{2016ZhouMobility}.
Figure~\ref{fig:diagnostics_1d}(b) shows the contribution of phonons with larger momentum, which are relevant for intervalley scattering.
For the linear conductivity, the momentum-resolved contribution is always negative, indicating that the scattering dissipates the current and enhances the resistivity.
In contrast, for the NLH effect, large-$q$ phonons with momentum slightly smaller than the intervalley distance enhance the NLH conductivity.
This result is consistent with our interpretation based on valley polarization.
Phonons with momentum smaller than the intervalley distance increase the anisotropy of the scattering (by increasing $\gamma_2$ in Fig.~\ref{fig:SM_scat_model} where $\gamma_2 > \gamma_1$), thus enhancing the valley polarization and the NLH conductivity.
Phonons with momentum larger than the intervalley distance tend to decrease the anisotropy of the scattering (by increasing $\gamma_1$ in Fig.~\ref{fig:SM_scat_model}) and thus suppress the NLH conductivity.

Figure~\ref{fig:diagnostics_gamma} shows the momentum-resolved conductivity for $q$ points near $\Gamma$.
For the inter-minivalley scattering, it is more difficult to analyze the role of each scattering vector, since a single scattering vector $\mb{q}$ contributes to many different scatterings $\mb{k}\to\mb{k+q}$.
Nevertheless, we find that the momentum-resolved contributions to linear conductivity and NLH conductivity show a large difference near the inter-minivalley distance, implying the importance of inter-minivalley scatterings on  enhancing the NLH conductivity.

\section{Scattering-induced valley polarizability} \label{sec:SM_valley_pol}
In order to demonstrate that the scattering-induced valley polarization is common in multi-valley semiconductors, we define and compute the valley polarizability.
Although here we consider only the ordinary valleys, not the minivalleys, the same analysis can be applied to the minivalleys if they are present.
For a given valley $V$ centered around the $k$ point $\mb{k}_V$, we define the valley polarization as
\begin{equation} \label{eq:valley_polarization_def}
    \delta n_V = \frac{1}{V_{\rm cell} \, N_k} \sum_{\mb{k} \in V} \sum_n \df_\nk.
\end{equation}
We define the valley polarizability $\mb{\alpha}_V$ as the valley polarization per current $\mb{J}$:
\begin{equation} \label{eq:valley_polarizability_def}
    \delta n_V = \sum_a \alpha_V^a J^a\,.
\end{equation}
For the valleys to be well-defined, different valleys must not overlap with each other and the nonequilibrium electron occupation $\df$ outside the valleys must be negligible.

In nonmagnetic semiconductors, one requirement for nonzero valley polarizability is that the valley is not at a time-reversal invariant momentum:
\begin{equation} \label{eq:valley_symmetry_time_reversal}
    \mb{\alpha}_V = \mb{0} \text{ if } \mb{k}_V = -\mb{k}_{V} + \mb{G}\,,
\end{equation}
where ${\bf G}$ is a reciprocal lattice vector.
This condition is required because the scattering-induced valley polarization transfers electrons from a valley to its time-reversal pair.
Formally, if $\mb{k}_V$ is a time-reversal invariant momentum, the corresponding valley polarization $\delta n_V$ should remain unchanged under time reversal, which reverses the sign of $\mb{J}$. Thus, \Eq{eq:valley_polarizability_def} gives $\delta n_V = -\delta n_V = 0$.

Also, if a valley $V$ is mapped to a valley $V'$ (which may or may not be indentical to $V$) by a spatial symmetry operation $s$ that maps a vector $\mb{k}$ to $s(\mb{k})$, the valley polarizibility should have the same symmetry:
\begin{equation}
    \mb{\alpha}_{V'} = s (\mb{\alpha}_V) \text{ if } \mb{k}_{V'} = s (\mb{k}_{V}) + \mb{G}\,.
\end{equation}
The reason is that the valley polarization at $V'$ under a current $s(\mb{J})$ should equal the valley polarization at $V$ under a current $\mb{J}$:
\begin{equation}
    \mb{\alpha}_{V'} \cdot s(\mb{J})
    = \mb{\alpha}_V \cdot \mb{J}\,.
\end{equation}
This condition forbids scattering-induced valley polarization in pristine graphene due to the C$_3$ symmetry~\cite{2021YingValleyPol}, because no nonzero in-plane vector can satisfy $\mathrm{C}_3\, \mb{\alpha}_K = \mb{\alpha}_K$.

Having an inversion symmetry is compatible with nonzero valley polarization if the above conditions are satisfied.
Consider an inversion-symmetric material in which inversion maps a valley $V$ to a different valley $V'$ satisfying $\mb{k}_{V'}=-\mb{k}_V+\mb{G}$.
Then, the constraint the inversion symmetry imposes on the valley polarization is $\mb{\alpha}_{V'}= - \mb{\alpha}_{V}$, and this value can be nonzero.
For an electric current $\mb{J}$, the valley polarizations read $\delta n_V = -\delta n_{V'} = \mb{\alpha}_V \cdot \mb{J}$.
If the current direction is inverted, the valley polarizations are also inverted: $\delta n_V = -\delta n_{V'} = -\mb{\alpha}_V \cdot \mb{J}$.

In the CRTA, the valley polarizability is always negligible because the number of carriers within each valley is almost conserved:
\begin{align}
    \delta n_V^{\rm CRTA}
    &\propto \tau\,\mb{E} \cdot \int_{V} d\mb{k} \mb{\nabla} f(\varepsilon_\nk) \nnnl
    &= \tau\,\mb{E} \cdot \int_{\partial V} \mb{dS} f(\varepsilon_\nk) \nnnl
    &\approx 0,
\end{align}
where $\partial V$ is the surface integral for the boundary of the valley.
The surface integral vanishes because $f(\varepsilon_\nk)$ is, to a good approximation, a constant at the boundary of the valley: 1 for the hole-doped case and 0 for the electron-doped case.

\begin{figure}[tb]
\centering
\includegraphics[width=1.0\columnwidth]{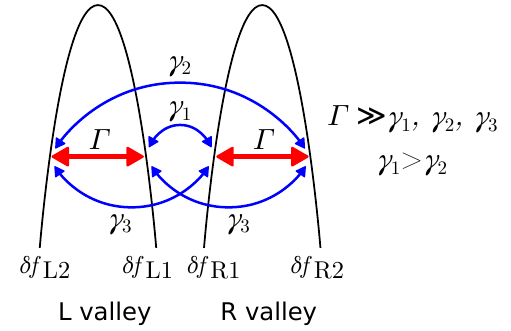}
\caption{
A four-state model for valley polarization.
}
\label{fig:SM_scat_model}
\end{figure}

To study how a finite valley polarization can emerge, let us consider the four-state model shown in Fig.~\ref{fig:SM_scat_model}.
This model provides a minimal description of the scattering-induced valley polarization.
It is inspired by the model of Ref.~\cite{2021YingValleyPol}, which has two circular Fermi pockets and scattering matrix elements that vary sinusoidally with the electron wavevector direction and was used to describe the scattering-induced valley polarization switching of twisted bilayer graphene.

The BTE of the four-state model reads
\begin{align} \label{eq:scat_model_bte}
    &\begin{pmatrix}
    \Gamma + \gamma_2 + \gamma_{3} & -\Gamma & -\gamma_3 & -\gamma_2 \\
    -\Gamma & \Gamma + \gamma_1 + \gamma_{3} & -\gamma_1 & -\gamma_3 \\
    -\gamma_3 & -\gamma_1 & \Gamma + \gamma_1 + \gamma_{3} & -\Gamma \\
    -\gamma_2 & -\gamma_3 & -\Gamma & \Gamma + \gamma_2 + \gamma_{3}
    \end{pmatrix} \nnnl
    \times &\begin{pmatrix}
    \df_{\rm L2} \\
    \df_{\rm L1} \\
    \df_{\rm R1} \\
    \df_{\rm R2}
    \end{pmatrix}
    = \begin{pmatrix}
    c \\
    -c \\
    c \\
    -c
    \end{pmatrix}.
\end{align}
Here, $c$ is a constant proportional to the band velocity and the external field.
Setting $c>0$ denotes a situation where the electrons move to the right and the current flows to the left.
This $4\times4$ scattering matrix has rank 3 due to charge conservation: the sum of each column is 0.
To find the physical stationary-state solution, we need to impose charge conservation
\begin{equation}
    \df_{\rm L1} + \df_{\rm L2} + \df_{\rm R1} + \df_{\rm R2} = 0.
\end{equation}
Let us assume that the intravalley scattering rate $\Gamma$ is much higher than the intervalley scattering rates $\gamma_1$, $\gamma_2$, and $\gamma_3$.
Then, the stationary-state solution of the BTE is
\begin{align}
    &\df_{\rm L1} = -\df_{\rm R1} \approx -\frac{\gamma_2 + \gamma_3}{\Gamma(\gamma_1 + \gamma_2 + 2\gamma_3)} c \nnnl
    &\df_{\rm L2} = -\df_{\rm R2} \approx \frac{\gamma_1 + \gamma_3}{\Gamma(\gamma_1 + \gamma_2 + 2\gamma_3)} c\,.
\end{align}
The corresponding valley polarization $\delta n = \df_{\rm L1} + \df_{\rm L2} - \df_{\rm R1} - \df_{\rm R2}$ is
\begin{align}
    \delta n \approx \frac{2(\gamma_1 - \gamma_2)}{\Gamma(\gamma_1 + \gamma_2 + 2\gamma_3)} c\,.
\end{align}
Thus, if the intervalley scattering rate is anisotropic, i.\,e.\,, $\gamma_1 \neq \gamma_2$, the electric field can generate a nonzero valley polarization.

\begin{figure*}[tb]
\centering
\includegraphics[width=1.0\textwidth]{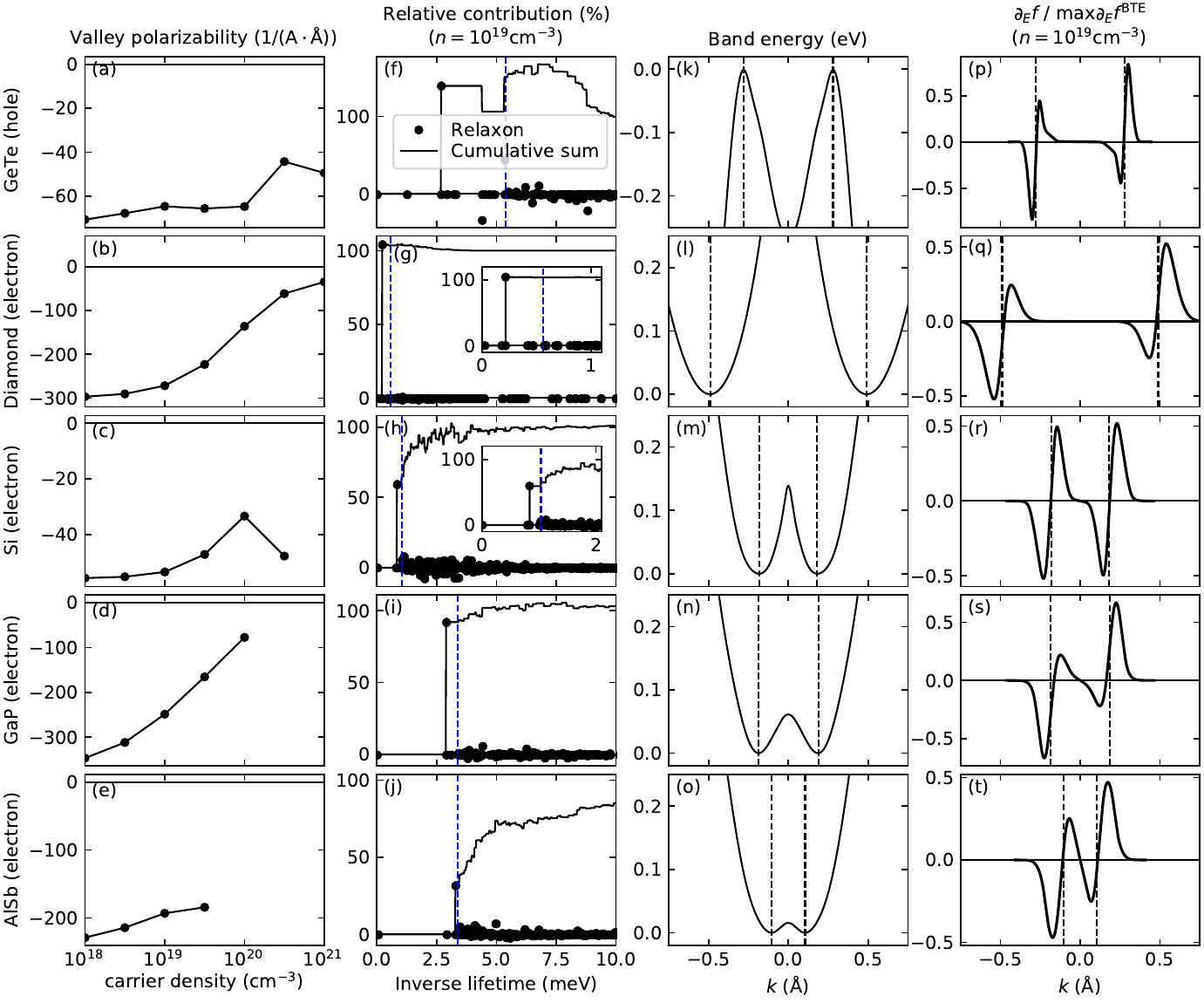}
\caption{
(a-e) Valley polarizability, (f-j) contribution of each relaxon to the valley polarizability, (k-o) band structure, and (p-t) electron occupation for five semiconductors at 300~K.
For the cubic semiconductors, the valley polarizabilities are shown for carrier densities low enough such that the chemical potential is below the X point energy and the CBM valleys are separated.
The dashed vertical lines in (k-t) indicate the position of the band extrema.
The $k$-point paths used in (k-t) are the one-dimensional cut shown in Fig.~\ref{fig:relaxon_dist}(a) for GeTe (k, p), and the $\mathrm{\Gamma}$-X-$\mathrm{\Gamma}$ path (a straight line in momentum space, parallel to the electric field) with the X point at $k=0$ for other systems (l-o, q-t).
}
\label{fig:valley_pol}
\end{figure*}

Finally, we compute the valley polarizability from first principles for the hole-doped GeTe and four electron-doped cubic semiconductors, diamond, silicon, GaP, and AlSb, which have conduction-band valleys around $k$ points that are not time-reversal invariant momenta.
For the cubic semiconductors, the valley polarizability vector is parallel to the valley momentum due to cubic symmetry: $\mb{\alpha}_V = \alpha \mb{k}_V / \abs{\mb{k}_V}$.
For GeTe, the valley polarizability is isotropic only within the $xy$ plane:
\begin{equation}
    \mb{\alpha}_V = \alpha_{\parallel}\, \mb{k}_{V, \parallel}  \,/\, \abs{\mb{k}_{V, \parallel}} + (0, 0, \alpha_{\perp})\, k_{V,z} \,/\, \abs{k_{V,z}}\,,
\end{equation}
where $\mb{k}_V = (\mb{k}_{V,\parallel}, k_{V,z})$.
For GeTe, we only compute the in-plane component $\alpha_{\parallel}$.

Figures~\ref{fig:valley_pol}(a-e) show the valley polarizability of the five materials.
We find that the valley polarizability is of the same order of magnitude in all cases, demonstrating that valley polarization generally occurs in mutivalley semiconductors, regardless of the presence of polar phonons or spin-orbit coupling.
From the relaxon decomposition in Figs.~\ref{fig:valley_pol}(f-j), we see that the long-lived relaxons make a significant and often dominant contribution to the valley polarizability.
Figures~\ref{fig:valley_pol}(k-o) show the relevant band structures, and Figs.~\ref{fig:valley_pol}(p-t) show the corresponding steady-state electron occupation obtained by solving the BTE.

This valley polarization occurs in the linear response regime and describes a transfer of electrons between a valley and its time-reversal pair.
This phenomenon is different from the polarization between valleys on different crystal axes under high electric fields studied in Refs.~\cite{2013IsbergValley,2021MaliyovHighField}.
We note that while Ref.~\cite{2021MaliyovHighField} states that the six valleys of silicon are equally occupied in the linear-response regime,
our analysis shows that valley polarization between the time-reversal pairs occurs in silicon already at the linear order.
Also, the valley polarization proposed in this work occurs for electric fields along any direction.
For example, if the electric field is applied along the [111] direction, the three valleys along the positive [100], [010], [001] directions will be less occupied than their inversion partners.

\section{Computational details} \label{sec:comp_details}

\begin{figure}[tb]
\centering
\includegraphics[width=0.99\linewidth]{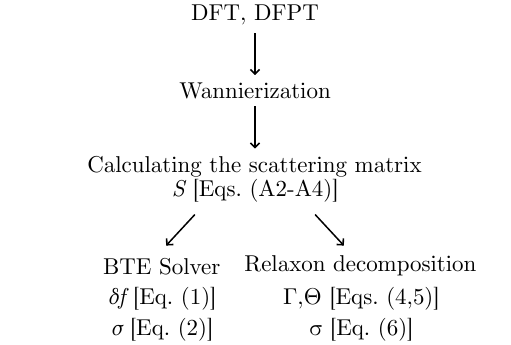}
\caption{
Workflow of the calculation.
}
\label{fig:workflow}
\end{figure}

Here, we discuss the details of the calculation performed in this work.
Figure~\ref{fig:workflow} summarizes the workflow of the calculation.

We used the \qe\ package~\cite{2017GiannozziQE} for the density functional theory and density functional perturbation theory calculations.
We used a 20$\times$20$\times$20 unshifted $k$-point grid, a kinetic energy cutoff of 80~Ry,
fully relativistic norm-conserving pseudopotentials~\cite{2013HamannONCVPSP} from the PseudoDojo library (v0.4)~\cite{2018VanSettenPseudoDojo}, and the Perdew-Burke-Ernzerhof (PBE) exchange-correlation functional~\cite{1996PerdewPBE}.
The lattice parameters and atomic coordinates were optimized until the stresses and forces were less than $1.0 \times 10^{-6}$~Ry/Bohr$^3$ and $1.0 \times 10^{-4}$~Ry/Bohr, respectively.
The optimized hexagonal lattice parameters, $a$=4.233~\AA, and $c$=10.921~\AA, and the Te shift $\tau$=0.34~\AA\ are slightly overestimated than the experimental values, $a$=4.16~\AA, $c$=10.67~\AA, and $\tau$=0.31~\AA~\cite{2016KrempaskyGeTe}.

We used the \wannier~\cite{2020PizziWannier90} and EPW~\cite{2016PonceEPW} codes to construct the Wannier functions and compute the real-space matrix elements.
The Brillouin zone was sampled with 12$\times$12$\times$12 and 6$\times$6$\times$6 grids for electrons and phonons, respectively.
We constructed 16 Wannier functions using the atomic $s$ and $p$ orbitals as initial guesses.
To preserve the crystal symmetries, we performed only the disentanglement step and not the maximal localization step.
We set the upper bound of the frozen (inner) and disentanglement (outer) windows at 3.3~eV and 7.7~eV above the valence band maximum, respectively.

We performed the Wannier interpolation, solved the BTE, and computed the transport coefficients using an in-house developed code \EPIjl, which is written in the Julia programming language~\cite{2017BezansonJulia}.
Details of the implementation will be discussed elsewhere.
For the solution of the BTE, we included only the states with energy in the range $[E_{\rm VBM} - 0.4~\mathrm{eV},\, E_{\rm VBM}]$, with $E_{\rm VBM}$ the valence-band maximum energy.
We sampled the Brillouin zone using unshifted 80$\times$80$\times$80 $k$- and $q$-point grids, and sampled only the irreducible Brillouin zone for the $k$ points utilizing the crystal and time-reversal symmetries.
To calculate the relaxons, we unfolded the scattering matrix from the irreducible Brillouin zone to the full Brillouin zone.
We solved the linear Boltzmann transport equations using the GMRES (generalized minimum residual) method as implemented in the \texttt{IterativeSolvers.jl} package~\cite{JuliaIterativeSolvers}.
A fixed numerical Gaussian smearing of 15~meV was applied to the energy-conserving delta functions in \myeqref{eq:scattering_mel}.
We ignored scattering processes that violate the total energy conservation by more than 2.5 times the smearing parameter.
Due to the higher computational cost of the relaxon calculation, we used a narrower window $[E_{\rm VBM} - 0.2~\mathrm{eV},\, E_{\rm VBM}]$, which is accurate enough for carrier densities below 10$^{20}$~cm$^{-3}$.
We used 100$\times$100$\times$100 $k$- and $q$-point grids to obtain the results shown in Figs.~\ref{fig:relaxon_decomp} and \ref{fig:relaxon_dist}.
One-dimensional plots of the relaxon eigenvector, velocity, and Berry curvature in Figs.~\ref{fig:relaxon_dist} were obtained by cubic spline interpolation.

To calculate the band velocity, we directly Wannier interpolated the velocity matrix elements, as done in Refs.~\cite{2021PonceMobility,2022LihmSpinPhoto}.
To calculate the Berry curvature, we used Eq.~(A1) of Ref.~\cite{2006WangAHC}.
We regularized the denominator including intermediate states to avoid divergence due to a vanishing denominator:
\begin{align}
    \mb{\Omega}_\nk(\omega) = -\sum_m &\Re \frac{(\varepsilon_\mk - \varepsilon_\nk)^2}{(\varepsilon_\mk - \varepsilon_\nk)^2 - (\hbar \omega + i \eta)^2} \nnnl
    &\times \Im (\mb{A}_{nm\mb{k}} \times \mb{A}_{mn\mb{k}}).
\end{align}
The regularization parameter was set to $\eta = 15~\mathrm{meV}$.
The frequency factor and the regularization parameter were only included in the second and third terms of Eq.~(A1) of Ref.~\cite{2006WangAHC}.
The frequency dependence of the first term can be safely neglected since it describes the coupling between a low-energy state near the Fermi level and a high-energy state outside the Wannierization window, so that $\abs{\varepsilon_\mk - \varepsilon_\nk} \gg \hbar \omega, \eta$.

Doping was treated using the rigid-band approximation.
The screening of the electron-phonon coupling from the free carriers was included~\cite{2017Verdi}.
The momentum- and frequency-dependent dielectric function $\epsilon(\omega, \mb{q})$ was calculated using the random phase approximation~\cite{MahanBook} and the material-specific electronic structure:
\begin{equation} \label{eq:screening_1}
    \chi_0(\omega, \mb{q}) = \frac{1}{V_{\rm cell}\, N_k} \sum_{\mb{k}} \sum_{n,m = 1}^{N_{\rm W}} \frac{(f_\nk - f_\mkq)\abs{M_{mn}(\mb{k},\mb{q})}^2}{\varepsilon_\nk - \varepsilon_\mkq + \hbar \omega + i\eta}
\end{equation}
and
\begin{equation}
    \epsilon(\omega, \mb{q}) = 1 - \frac{4\pi e^2}{\mb{q} \cdot \mb{\epsilon^{\infty}} \cdot \mb{q}} \chi_0(\omega, \mb{q}) \, ,
\end{equation}
where
\begin{equation} \label{eq:screening_3}
    M_{mn}(\mb{k},\mb{q}) = [U^\dagger(\mb{k+q}) \, U(\mb{k})]_{mn}
\end{equation}
is the overlap of the eigenvectors at $\mb{k}$ and $\mb{k+q}$ in the Wannier basis, and $\mb{\epsilon^{\infty}}$ the infinite-frequency susceptibility tensor of the undoped system.
This method is similar to Eq.~(8) of Ref.~\cite{2022MachedaDoping}, with the differences that in our work (i) the frequency dependence of the dielectric function is taken into account, and (ii) the matrix element is approximated as $\braket{u_{m\mb{k+q}}}{u_{n\mb{k}}} \approx M_{mn}(\mb{k},\mb{q})$ instead of $\braket{u_{m\mb{k+q}}}{u_{n\mb{k}}} \approx \delta_{mn}$.
We used $\eta$=15~meV which was chosen based on the calculated electron quasiparticle lifetimes.
We screened only the long-range Fr\"ohlich part, not the Wannier-interpolated short-range part, because the free-carrier screening dominantly acts on the long-range (small-$q$) component of the potential, and the dielectric function $\epsilon(\omega, \mb{q})$ calculated using Eqs.~(\ref{eq:screening_1}-\ref{eq:screening_3}) is intended to work only in that limit.
Concretely, we used~\cite{2022MachedaDoping}
\begin{equation}
    g^{\rm scr.}_{mn\nu}(\mb{k},\mb{q}) = g^{\rm short-range}_{mn\nu}(\mb{k},\mb{q}) + g^{\rm Fr\ddot{o}hlich}_{mn\nu}(\mb{k},\mb{q}) / \epsilon(\omega_\qnu, \mb{q}).
\end{equation}

The quadrupole contribution to the phonon dynamical matrix~\cite{2019RoyoQuadrupole} and the electron-phonon coupling~\cite{2020BruninQuadrupole,2020JhalaniQuadrupole} was not included.

For the calculation of cubic semiconductors, diamond, silicon, GaP, and AlSb, we used the calculation parameters used in Ref.~\cite{2021PonceMobility} unless otherwise specified.
For the cubic semiconductors, we neglected spin-orbit coupling and used scalar-relativistic pseudopotentials.
The Wannier functions were obtained with a 12$\times$12$\times$12 grid and a 6$\times$6$\times$6 grid for electrons and phonons, respectively.
To calculate the valley polarization with the BTE, we sampled the $k$ points with an $N_k\times N_k \times N_k$ grid with $N_k=150$ and used 15~meV smearing for the energy-conserving delta function.
For the relaxon calculation we used $N_k=120$ for silicon and AlSb, and $N_k=100$ for diamond and GaP after a convergence study.

The Brillouin zone plots in Fig.~\ref{fig:relaxon_dist}(a, b) were generated using \texttt{Brillouin.jl}~\cite{JuliaBrillouin}.
Cubic spline interpolation was performed using \texttt{Interpolations.jl}~\cite{JuliaInterpolations}.

We note that Ref.~\cite{2019AskarpourGeTe} also provides a first-principles calculation of the mobility of GeTe considering electron-phonon scattering using the momentum relaxation time approximation.
The linear mobility of GeTe reported there is about 50\% smaller than that of our calculation.
A possible explanation is that the calculation of the band velocity neglecting the contribution of the nonlocal pseudopotential in Ref.~\cite{2019AskarpourGeTe} led to an underestimation of the band velocity and mobility.
It has been shown in Ref.~\cite{2021PonceMobility} that such an approximation can lead to a sizable relative error in the mobility, ranging from $-$71\% to $+$37\%, depending on the material.
This explanation is supported by the fact that the scattering rates of Ref.~\cite{2019AskarpourGeTe} are in good agreement with our results, while the mean-free paths, which are the band velocities divided by the scattering rates, are 2 to 3 times smaller than our result.

\FloatBarrier 

\makeatletter\@input{xx.tex}\makeatother
\bibliography{main}